\documentclass[pra,twocolumn,floatfix,superscriptaddress,amsmath,amssymb,citeautoscript,aps,longbibliography]{revtex4-2}

\usepackage{amsfonts,amsmath,amssymb,color,times,graphicx}
\usepackage[colorlinks=true, breaklinks=true, linkcolor=blue, citecolor=blue, urlcolor=blue]{hyperref}

\usepackage{blindtext}
\usepackage{graphicx}
\usepackage{dcolumn}
\usepackage{bm}
\usepackage{tabularx}
\usepackage{lineno}
\usepackage{eqnarray,amsmath}
\usepackage{lipsum}
\newcommand{\YAO}{YbAlO$_3$}
\newcommand{\DSO}{DyScO$_3$}
\newcommand{\TSO}{TbScO$_3$}

\usepackage{color}
\definecolor{magenta}{cmyk}{ 0, 1, 0,0}
\usepackage[dvipsnames]{xcolor}

\begin{document}
\title{Quasi One-Dimensional Ising-like Antiferromagnetism in the Rare-earth Perovskite Oxide TbScO$_3$ }
\date{\today}
\author{Nan Zhao}
\affiliation{Department of Physics, Southern University of Science and Technology, Shenzhen 518055, China}

\author{Jieming Sheng}
\thanks{Corresponding author: shengjm@sustech.edu.cn}
\affiliation{Department of Physics, Southern University of Science and Technology, Shenzhen 518055, China}
\affiliation{Academy for Advanced Interdisciplinary Studies, Southern University of Science and Technology, Shenzhen, 518055, China}

\author{Jinchen Wang}
\affiliation{Department of Physics, Renmin University of China, Beijing 100872, China}

\author{Han Ge}
\affiliation{Department of Physics, Southern University of Science and Technology, Shenzhen 518055, China}

\author{Tiantian Li}
\affiliation{Department of Physics, Southern University of Science and Technology, Shenzhen 518055, China}

\author{Jiong Yang}
\affiliation{Department of Chemistry, Southern University of Science and Technology, Shenzhen 518055, China}

\author{Shanmin Wang}
\affiliation{Department of Physics, Southern University of Science and Technology, Shenzhen 518055, China}
\affiliation{Quantum Science Center of Guangdong-Hong Kong-Macao Greater Bay Area (Guangdong), Shenzhen 518045, China}

\author{Ping Miao}
\affiliation{Institute of High Energy Physics, Chinese Academy of Sciences, Beijing 100049, China}
\affiliation{Spallation Neutron Source Science Center, Dongguan 523803, China}

\author{Hua He}
\affiliation{School of Chemical Science, University of Chinese Academy of Sciences (UCAS), Beijing 100190, China}

\author{Xin Tong}
\affiliation{Institute of High Energy Physics, Chinese Academy of Sciences (CAS), Beijing 100049, China}
\affiliation{China Spallation Neutron Source (CSNS), Dongguan 523803, China}

\author{Wei Bao}
\affiliation{Department of Physics, City University of Hong Kong, Kowloon, Hong Kong}

\author{Er-Jia Guo}
\affiliation{Beijing National Laboratory for Condensed Matter Physics and Institute of Physics,Chinese Academy of Sciences, Beijing 100190, China}
\affiliation{University of Chinese Academy of Sciences, Beijing 100049, China}
\affiliation{Songshan Lake Materials Laboratory, Dongguan, Guangdong 523808, China}

\author{Richard Mole}
\affiliation{Australian Nuclear Science and Technology Organisation, Lucas Heights, New South Wales 2234, Australia}

\author{Dehong Yu}
\affiliation{Australian Nuclear Science and Technology Organisation, Lucas Heights, New South Wales 2234, Australia}

\author{Andrey A Podlesnyak}
\affiliation{Neutron Scattering Division, Oak Ridge National Laboratory, Oak Ridge, TN 37831, USA}

\author{Liusuo Wu}
\thanks{Corresponding author: wuls@sustech.edu.cn}
\affiliation{Department of Physics, Southern University of Science and Technology, Shenzhen 518055, China}
\affiliation{Quantum Science Center of Guangdong-Hong Kong-Macao Greater Bay Area (Guangdong), Shenzhen 518045, China}
\affiliation{Shenzhen Key Laboratory of Advanced Quantum Functional Materials and Devices, Southern University of Science and Technology, Shenzhen 518055, China}

\begin{abstract}
The rare-earth perovskite~\TSO~has been widely used as a substrate for the growth of epitaxial ferroelectric and multiferroic thin films, while its detailed low-temperature magnetic properties were rarely reported. In this paper, we performed detailed magnetization, specific heat and single crystal neutron scattering measurements, along with the crystalline electric field calculations to study the low-temperature magnetic properties of~\TSO. All our results suggest the magnetic Tb$^{3+}$ has an Ising-like pseudo-doublet ground state at low temperatures. Due to the constrain of local point symmetry, these Tb$^{3+}$ Ising moments are confined in the $ab$ plane with a tilt angle of $\varphi = \pm48^{\mathrm{o}}$ to the $a$ axis. In zero field, the system undergoes an antiferromagnetic phase transition at $T_{\mathrm{N}}=2.53$~K, and forms a $G_xA_y$ noncollinear magnetic structure below $T_{\mathrm{N}}$. We find the dipole-dipole interactions play an important role to determine the magnetic ground state, which are also responsible for the quasi-one-dimensional magnetism in \TSO. The significant anisotropic diffuse scatterings further confirm the quasi-one-dimensional magnetism along the $c$ axis. The magnetic phase diagram with the field along the easy $b$ axis is well established. In addition to the $G_xA_y$ antiferromagnetic state, there is an exotic field-induced phase emerged near the critical field $B_{\mathrm{c}}\simeq0.7$~T, where three-dimensional magnetic order is suppressed but strong one-dimensional correlations may still exist.
\end{abstract}

\maketitle

\section{Introduction}
\maketitle
The family of orthorhombic rare-earth perovskite oxides $RM$O$_3$ (where $R$ is a $4f$ rare-earth ion, $M$ is a $3d$ transition-metal ion) attract continued attention due to their numerous intriguing physical phenomena and potential applications. Remarkable examples include ferroelectric and multiferroic properties \cite{Kimura2003,Goto2004}, magneto-optical effects \cite{Kimel2004,Kimel2005}, magnetocaloric effects \cite{Wuyaodong2019, Jia2019}, exotic quantum spin states \cite{Nikitin2018,Wu2019A,Podlesnyak2021}, and complex spin reorientation transitions \cite{Gorodetsky1968,Belov1976} induced by temperature and magnetic field. Many of these interesting properties arise from the peculiarity of $3d$ and $4f$ magnetic sublattices and the diversity of exchange interactions between them. In general, for $RM$O$_3$ with a $3d$ magnetic ion, the strong exchange coupling between the neighboring $M$ magnetic ions could induce a robust canted antiferromagnetic (AFM) order with a weak ferromagnetic component at several hundreds of degrees Kelvin \cite{Yamaguchi1973,Shapiro1974,Hahn2014}. As the temperature decreases, the $R-M$ coupling becomes more significant and induces spin-reorientation transitions around several dozen Kelvin \cite{White1969,Yamaguchi1974}. The strong coupling between $3d$ and $4f$ magnetic sublattices gives rise to a series of nontrivial magnetism, including exotic solitonic lattice \cite{Artyukhin2012}, gigantic magnetoelectric effect \cite{Tokunaga2008}, unconventional low-energy spin excitations \cite{Nikitin2018,Podlesnyak2021}. In contrast, due to the weak exchange coupling between the $4f$ moments, the rare-earth sublattice orders at much lower temperatures \cite{Wu2017,Wu2019A,Sheng2020}, usually below 10 K. From naive expectation, these rare-earth ions would behave rather classically, considering the large saturation moments and the strong anisotropy. Maybe partly due to this reason, the low-temperature ground states of these rare-earth sublattices remain poorly studied.

Recent work on the ytterbium (Yb) based perovskite compound YbAlO$_3$ reveals that the effective spin-1/2 state emerged when the doublet ground state is realized due to the crystalline electric field (CEF) splittings \cite{Wu2019B}. In addition, fractional spinon excitations and quantum critical Tomonaga-Luttinger liquid behavior have been observed \cite{Wu2019A}. Meanwhile, no inelastic excitation but strong diffuse scattering was observed in the isostructural DyScO$_3$ \cite{Wu2017}. These distinct phenomena stimulate us to explore more perovskite $RM$O$_3$ systems with different rare-earth sublattices.

The rare-earth scandate perovskites $R$ScO$_3$ are important members in the $RM$O$_3$ families. High-quality $R$ScO$_3$ single crystals have been widely used as the substrates for various epitaxial ferroelectric and multiferroic thin film growth. The effect of biaxial strain from the small lattice mismatch between substrates and thin films can dramatically change the properties of epitaxial thin films \cite{Schlom2007}. For example, the SrTiO$_3$ thin film exhibits strain-induced ferroelectricity at room temperature when it is grown on \DSO~ substrate \cite{Haeni2004}. In addition to the effects of strain, it was proposed that the intrinsic magnetism from $R$ScO$_3$ substrates may also play an important influence on the electronic properties of thin films, especially at low temperatures \cite{Lustikova2022}. Very recently, it was reported that the strong anisotropic magnetism in \DSO~substrate could induce anisotropic and hysteretic magnetoresistance in the epitaxial Sr$_{1-x}$La$_x$CuO$_2$ thin film \cite{Lustikova2022}. However, due to the low ordering temperature of $RM$O$_3$ compounds, the influences of the magnetism from $R$ScO$_3$ substrates to the epitaxial thin films are rarely discussed, and the detailed magnetic properties of $R$ScO$_3$ remain to be explored.

Here, we investigated the terbium (Tb) based rare-earth scandate perovskite TbScO$_3$~by magnetization, specific heat, and neutron scattering measurements. At low temperatures, an Ising-like pseudo-doublet ground state with strong single-ion anisotropy was realized for these magnetic Tb$^{3+}$ ions. These Tb$^{3+}$ moments are confined in the $ab$ plane and canted by $\pm48^{\mathrm{o}}$ with respect to the $a$ axis. Below $T_{\mathrm{N}}=2.53$~K, a $G_xA_y$ magnetic long-range order state is selected by the dipolar interactions. Similar to the iso-structure compounds DyScO$_3$~\cite{Wu2017} and YbAlO$_3$~\cite{Wu2019B}, the interactions along the $c$ axis are dominant, suggesting a quasi-one-dimensional antiferromagnetism in~\TSO. Consistently, strong anisotropic diffuse scattering was observed in the neutron scattering experiments, which further confirmed the one-dimensional AFM magnetism. Combining these experimental results, a detailed field-induced magnetic phase diagram was established.
We found a gradual suppression of the zero-field antiferromagnetic order with increasing field along the local easy axis. An additional field-induced phase emerged above the critical field $B_{\mathrm{c}}\simeq0.7$~T at temperatures below $\sim$1 K, which is beyond the mean-field expectations of classical Ising moments.
\begin{figure*}[t]
 \includegraphics[width=0.9\textwidth]{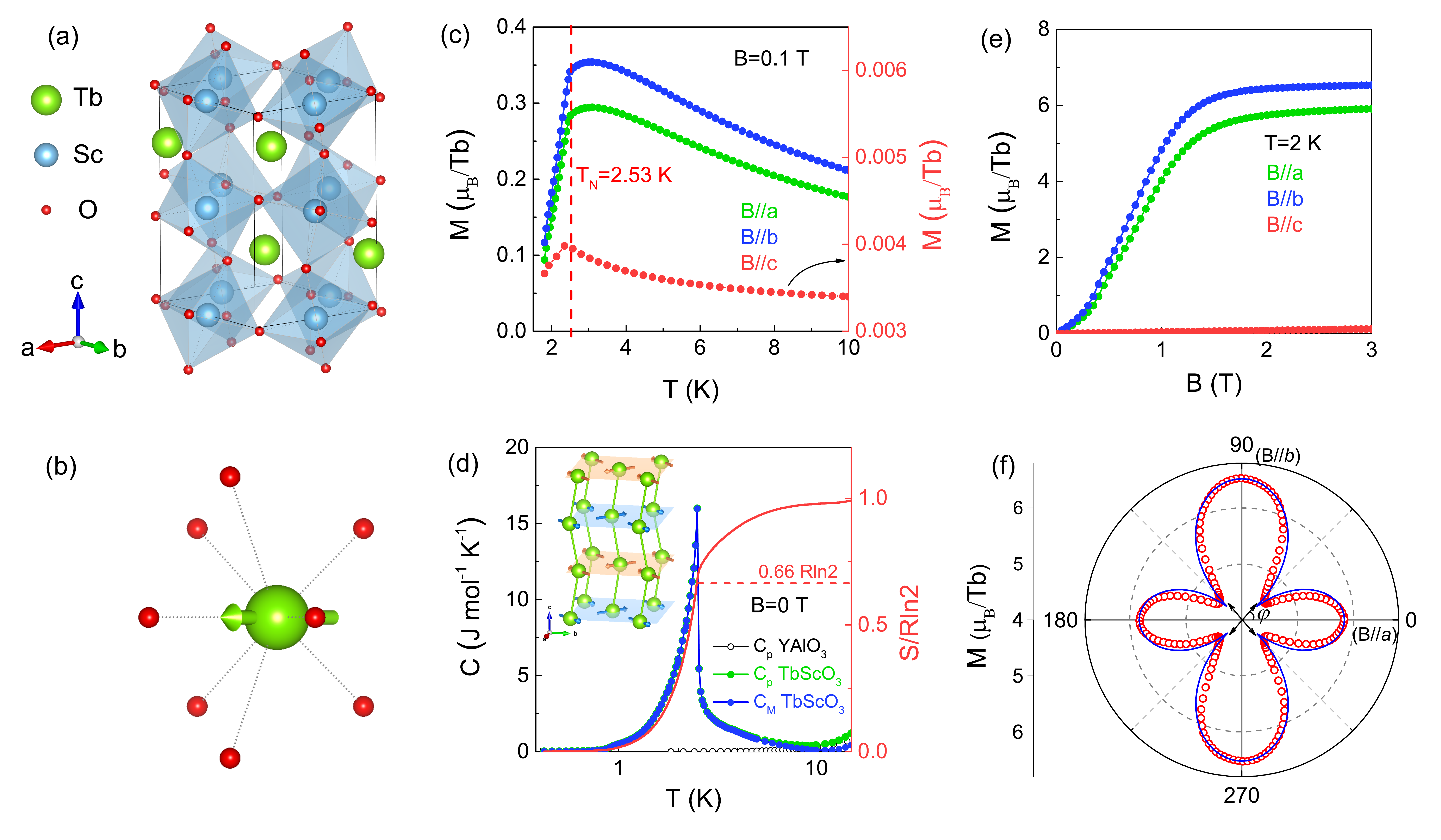}
    \caption{(a) The crystal structure of TbScO$_3$. (b) The local environment of Tb$^{3+}$ by considering the 8 nearest neighbors of oxygen atoms. (c) Temperature dependence of magnetization $M(T)$, measured at $B=0.1$~T, with magnetic field along the $a$ (green circles), $b$ (blue circles), $c$ (red circles)  directions, respectively. (d) The zero-field specific heat, $C_{\mathrm{p}}(T)$ (green solid circles), and magnetic specific heat, $C_{\mathrm{M}}(T)$ (blue solid circles) for \TSO. The black empty circles represent the specific heat of the non-magnetic YAlO$_3$. The red line is the integrated magnetic entropy of \TSO. (e) Field evolution of the magnetization $M(B)$, measured at $T=2$ K, with field along the three different principal crystallographic axes. (f) Angle dependence of the magnetization measured at $B=3$ T with applying field in the $ab$ plane.}\label{structure}
\end{figure*}
\section{Experimental Details}
Single crystals of~\TSO~were grown by the Czochralski technique~\cite{Uecker2008}, which are commercially available as substrates for thin film growth. For the magnetization measurements, two magnetometers were applied for different temperature ranges: (i) Quantum Design Magnetic Property Measurement System (MPMS) for the temperature range $T=1.8-300$~K; (ii) High-sensitive Hall sensor magnetometers~\cite{Hallsensor,Cavallini2004,Candini2006} for the lower temperature range $T=0.1-4.0$~K. The specific heat measurements were performed using the relaxation time method in a Quantum Design Physical Property Measurements System (PPMS) with a temperature range $T=1.8-300$~K. For the low temperature ($T=0.4-4.0$~K) measurements, a dilution refrigerator insert was used.

The single crystal neutron scattering experiments were carried out using the time-of-flight cold neutron spectrometer, PELICAN \cite{Yu2013}, with fixed incident energy $E_{\mathrm{i}}=3.71$~meV at the open pool Australian light-water multi-purpose reactor (OPAL) at Australian Nuclear Science and Technology Organization (ANSTO), and the multiplexing cold neutron spectrometer, BOYA~\cite{Wang}, at the China Advanced Research Reactor. The data were collected on a 600 mg single crystal~\TSO. The sample was aligned in the (H,0,L) scattering plane and cooled using a cryostat (3~K--300~K) at BOYA and a cryo-magnet (1.5 K - 800 K, maximum vertical field of 7 T)  at PELICAN.  The data were processed using the data analysis and visualization environment (DAVE) software package~\cite{Azuah2009}, and the Horace program~\cite{Ewings2016}.

\section{Results and Analysis}
\subsection{Crystal structure and pseudo-doublet ground state}
\TSO~crystallizes in a distorted orthorhombic perovskite structure with the $Pbnm$ (No.~62) space group. The lattice parameters refined from the single crystal X-ray diffraction measured at $T=100$ K are $a=5.4301(1)$~\AA, $b=5.7004(2)$~\AA, $c=7.8512(2)$~\AA\ and $\alpha=\beta=\gamma=90^{\rm \mathrm{o}}$, that is consistent with the previous report~\cite{Uecker2008}. As illustrated in Fig.~\ref{structure}(a), the Tb$^{3+}$ ions located at $4c$ Wyckoff sites of local point group $C_{\rm \mathrm{s}}$, are surrounded by eight distorted ScO$_6$ octahedrons. In this distorted environment, only one mirror plane normal to the $c$ axis is preserved, which constrains the Tb$^{3+}$ moment either along the $c$ axis or lies in the $ab$ plane. The CEF calculation based on the point-charge model was performed using the software package MCPHASE \cite{MCPHASE2004}. Eight nearest oxygen atoms around the Tb$^{3+}$ ion are included, and the 13-fold ($2J+1=13$) degenerate $J=6$ multiple states are split into separated singlet states [Fig.~\ref{structure}(b)]. Although Tb$^{3+}$ is a non-Kramers ion, the CEF calculations indicate a quasi-doublet ground state, with a very small energy gap (less than $0.001$ meV), and this ground quasi-doublet state is well separated from the next excited crystal field level (about 27.03 meV). Thus, we can treat the ground state as an effective $S_{\mathrm{eff}}=1/2$ state, similar to the case of the isostructural compounds DyScO$_3$~\cite{Wu2017} and YbAlO$_3$~\cite{Wu2019B}. Following the same strategy, as described for DyScO$_3$ \cite{Wu2017}, we find the Ising axis of the moment in the ab-plane with $\varphi = 46.2^\mathrm{o}$  titled away from the a-axis. This results in the ground state wave functions given by:
\begin{eqnarray*}
 |E_{0\pm}\rangle\ =\ &0.706 |-6\rangle \pm 0.706 |+6\rangle\\
     -&0.016 |-4\rangle \mp 0.016|+4\rangle\\
     +&0.028|-3\rangle\mp 0.028|+3\rangle\\
     +&0.005|-2\rangle  \pm 0.005 |+2\rangle\\
     -&0.006|-1\rangle \pm 0.006|+1\rangle.
\end{eqnarray*}
If we only keep the leading terms, this quasi-doublet ground state can be simplified as:
\begin{eqnarray*}
 |E_{0\pm}\rangle\ &\simeq\  &0.706 |-6\rangle \pm  0.706 |+6\rangle\\ &\simeq\ &\frac{1}{\sqrt{2}}(|-6\rangle \pm  |+6\rangle ).
\end{eqnarray*}
These results are just as expected from the irreducible representation analysis of the point symmetry $C_{\mathrm{s}}$~\cite{Valiev2021}. In zero field, this wave function indicates that the quasi-doublet ground state consists of the two closely spaced singlet states which are linear combinations of pure $|\pm6\rangle$ states, and thus zero magnetic moment is expected. With the external field applied along the local easy axis, these wave functions are then mixed with each other, and magnetic moments are induced~\cite{Holmes1971,Valiev2021}. In other words, the fact that these Tb moments can be easily polarized in small magnetic fields, also confirms that the ground state of~\TSO~is a quasi-doublet. The saturation moment and the calculated tilting angle are further verified with the magnetization measurements.
\begin{figure*}[ht!]
 \includegraphics[width=0.9\textwidth]{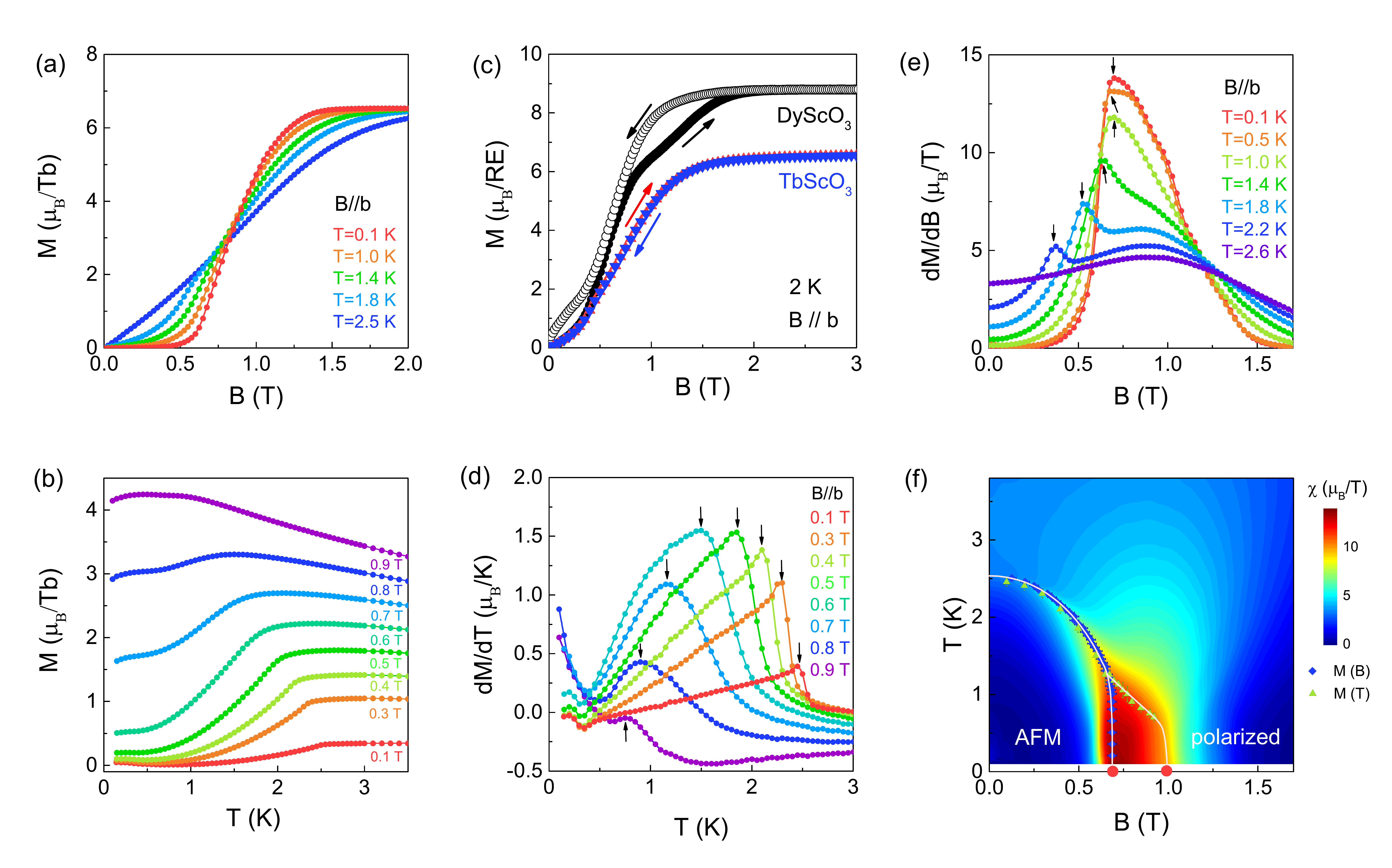}
    \caption{(a) Field evolution of the magnetization $M(B)$ measured at different temperatures with $B \parallel b$. (b) Temperature-dependent magnetization $M(T)$ measured at different fields with $B \parallel b$. (c) Field-dependent magnetization with field sweep up and sweep down measured at 2~K for DyScO$_3$~\cite{Wu2017} and TbScO$_3$. (d) The temperature deviation of magnetization $dM/dT$ measured at different magnetic fields. (e) Magnetic susceptibilities $dM/dB$ measured at different temperatures. (f) The contour plot of the magnetic phase diagram of \TSO, based on the magnetic susceptibilities $dM/dB$, with $B \parallel b$. }\label{magnetization}
\end{figure*}
\subsection{Magnetic Ground State}
To determine the ground state of \TSO, we performed the low temperature magnetization measurements. Shown in Fig.~\ref{structure}(c) is the temperature-dependent magnetization $M(T)$ in an external field of 0.1~T along the $a$ (green circles), $b$ (blue circles), and $c$ (red circles) directions, respectively. An AFM phase transition was identified at $T_{\mathrm{N}}=2.53$~K, as illustrated by the red dashed line in Fig.~\ref{structure}(c). Consistently, a sharp peak was observed in the zero-field specific heat $C_{\mathrm{p}}(T)$ (green solid circles) at $T=2.5$~K, as illustrated in Fig.~\ref{structure}(d), which further confirmed the AFM phase transition. The magnetic specific heat $C_{\mathrm{M}}(T)$ (blue solid circles) was obtained by subtracting the phonon contribution, estimated from the non-magnetic isostructural compound YAlO$_3$. The magnetic entropy was then calculated by integrating $C_{\mathrm{M}}/T$ from 400~mK to 30~K. A total entropy of about $R\mathrm{ln}2$ was obtained, consistent with the expectation of a pseudo-doublet ground state from the CEF calculations. It is interesting to note that, only about $66\%R\mathrm{ln}2$ entropy is released at $T_{\mathrm{N}}$, which indicates that partial moments are still fluctuating in the AFM ordered state.

Figure~\ref{structure}(e) shows the field evolution of the magnetization $M (B)$ measured at 2~K with field along the three principal crystallographic axes. Similar to the iso-structure compound TbAlO$_{3}$ \cite{Holmes1968}, strong magnetic anisotropy was observed between the $ab$ plane and the $c$ axis. The saturation moments were found to be 5.91~$\mu_{\mathrm{B}}$ and 6.53~$\mu_{\mathrm{B}}$ for $B \parallel a$ and $B \parallel b$, respectively. These values are nearly 50 times larger than the moment 0.12~$\mu_{\mathrm{B}}$ measured at 3~T along the $c$ axis, confirming that these Tb$^{3+}$ moments are lying in the $ab$ plane. To further determine the specific orientation of these moments, angle-dependent magnetization was measured in the $ab$ plane. The magnetization data (red empty circles) were collected at $T=2$~K and $B=3$~T, as shown in Fig.~\ref{structure}(f). Petal-like features were observed, as the two Ising sublattices were polarized in magnetic fields. Assuming a canted angle $\varphi$ between the Ising moments (black arrows in Fig.~\ref{structure}(f)) and the $a$ axis, the measured angular dependent magnetization can be described as:
\begin{equation}
  \label{Angular_M}
  M=\frac{M_{\mathrm{s}}}{2}(|\cos(\theta + \varphi)| + |\cos(\theta - \varphi)|),
\end{equation}
where $M_{\mathrm{s}}$ is the saturation moment, $\theta$ is the angle between the magnetic field and the $a$ axis. The best fit of the magnetization is shown as the blue solid line in Fig.~\ref{structure}(f), with $M_{\mathrm{s}}=8.8~\mu_{\mathrm{B}}$ and $\varphi=48^{\mathrm{o}}$. This tilting angle from the fitting of the angular dependent magnetization is very close to the value of 46.2$^{\rm o}$ obtained from the point charge calculations.

\begin{figure*}[ht!]
 \includegraphics[width=0.9\textwidth]{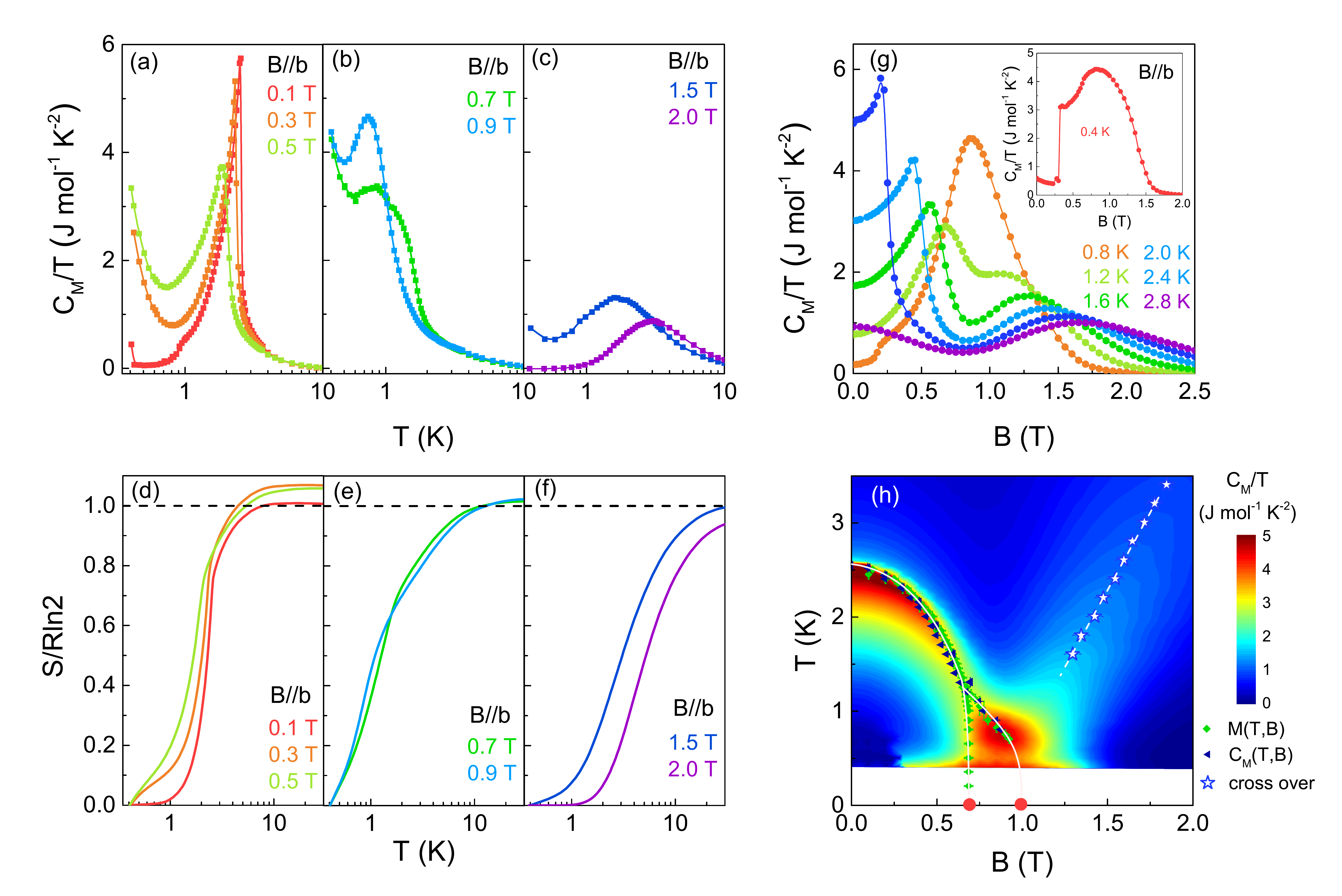}
    \caption{(a)-(c) Temperature dependence of specific heat $C_{M}/T$ of \TSO~in different fields, with $B \parallel b$. (d)-(f) Temperature-dependent magnetic entropy, normalized to $R$ln2, assuming doublet ground states, with $B \parallel b$. (g) Magnetic field dependence of specific heat $C_{M}/T$ of \TSO~in different temperatures, with $B \parallel b$. (h) The magnetic phase diagram of TbScO$_3$, based on the magnetic specific heat, $C_M/T$, with $B \parallel b$.}\label{HC}
\end{figure*}

\subsection{Magnetic dipolar interactions}
Since the $4f$ electrons are usually very localized, the exchange interactions between the neighboring rare-earth moments are relatively weak. On the other hand, due to the large local moments, the magnetic dipole-dipole interactions are expected to play an important role. For TbScO$_3$, four different magnetic structures ($A_xG_y, G_xA_y, F_xC_y,$ and $C_xF_y$) are allowed by the representation analysis, within the given crystal structure and the magnetic propagation vector $K=(0,0,0)$~\cite{Wu2017, Wu2019B}. It is found that the dipolar interactions select the zero-field magnetic ground state from these four possible spin configurations~\cite{Wu2017,Wu2019B}.

Here, we calculated the dipole-dipole energy for each of these four magnetic structures by fixing the Ising moment in the $ab$ plane with a canted angle $\varphi=\pm 48^{\mathrm{o}}$ to the $a$ axis:
\begin{equation}
 \label{Dipole_E}
  E_{\mathrm{dip}}=-\frac{\mu_0}{4\pi}\sum_{i}\frac{1}{{|\vec{r}|}^3} \cdot [3(\vec{m_0} \cdot {\hat{r}_i}) (\vec{m_i} \cdot {\hat{r}_i})-(\vec{m_0} \cdot {\vec{m_i}})],
\end{equation}
where $\mu_0$ is the vacuum permeability, $|\vec{r}|$ is the distance between two moments, and $\hat{r}_i=\vec{r}/|\vec{r}|$. For the calculation, ten neighboring Tb$^{3+}$ were considered in total, including eight near neighbors within a distance of 5.73~\AA\ in the $ab$ plane, and two neighbors at a distance of 4.02~\AA\ along the $c$ axis. We found that the influence was negligible for more distant atoms. The calculated dipole-dipole energies for each magnetic structures are $E_{\mathrm{dip}}({G_xA_y}) =-2.33$~K, $E_{\mathrm{dip}}(A_xG_y)=-1.47$~K, $E_{\mathrm{dip}}(C_xF_y)=0.40$~K and $E_{\mathrm{dip}}(F_xC_y)=1.26$~K. The spin configuration $G_xA_y$ [see inset of Fig.~\ref{structure}(d)] gains more energy than all the other three configurations, and the calculated dipole-dipole energy of -2.33 K is also comparable with the measured AFM transition temperature $T_{\mathrm{N}}=2.53$~K. In addition, we noticed that the dipolar interaction, $E^c_{\mathrm{dip}}=-1.37$~K, between the two nearest neighbors along the $c$ direction is larger than the in-plane dipolar interaction $E^{ab}_{\mathrm{dip}}=-0.42$~K, even with the four nearest  neighbors taken into consideration. This suggests that Tb$^{3+}$ ions form quasi-one-dimensional AFM spin chains along the $c$ axis (see inset of Fig.~\ref{structure}(d)) similar to the isostructural compounds TbAlO$_3$~\cite{Hufner1967}, \DSO~\cite{Wu2017} and \YAO~\cite{Wu2019A,Wu2019B}.

\begin{figure*}[ht!]
 \includegraphics[width=0.9\textwidth]{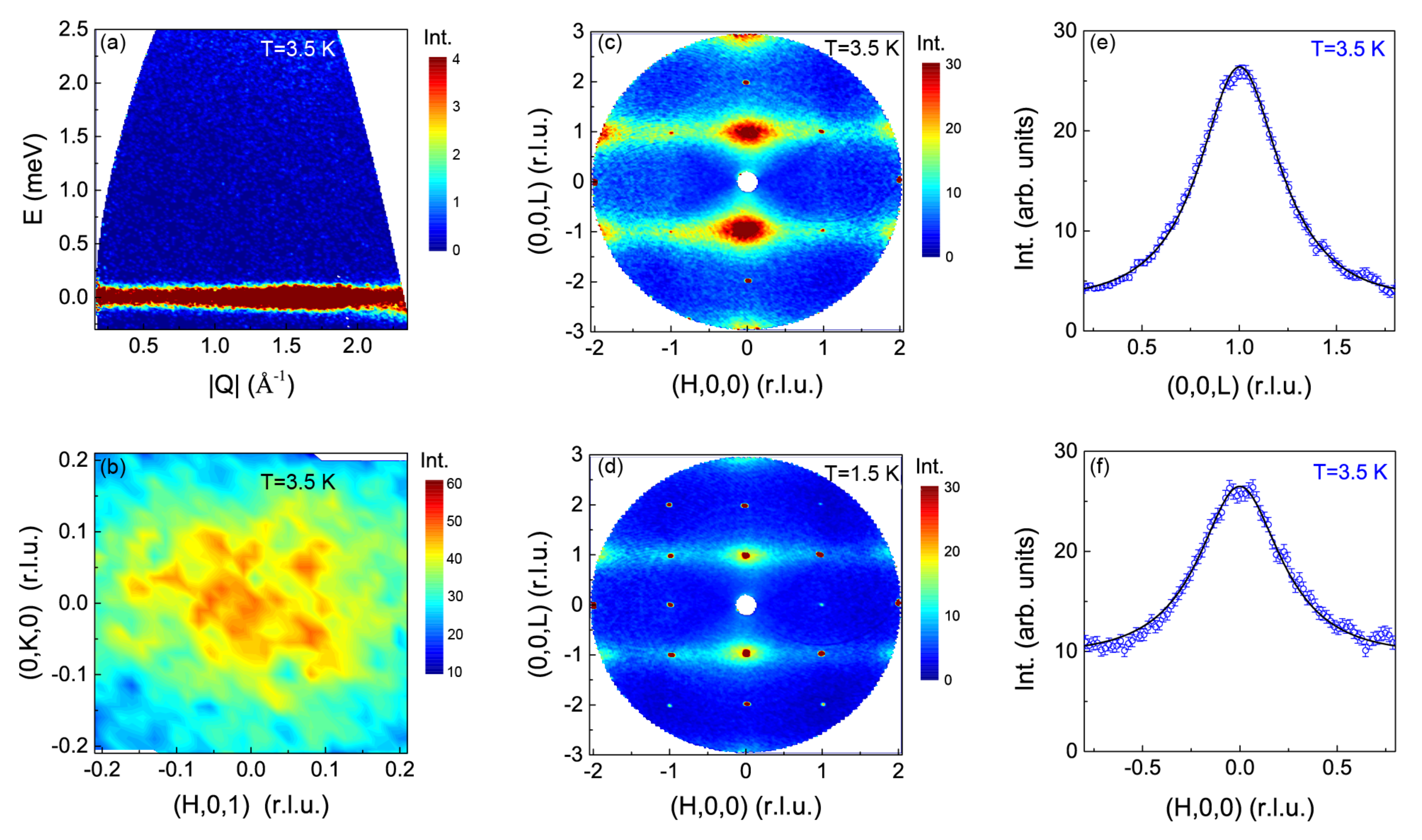}
    \caption{(a) Inelastic neutron scattering spectra of \TSO~measured at 3.5 K. (b) The diffuse scattering of \TSO~in the (H,K,0) scattering plane. The magnetic diffuse scattering of \TSO~in the (H,0,L) scattering plane with the integrated energy $E=[-0.1, 0.1]$ meV, measured at (c) $T=3.5$ K and (d) $T=1.5$ K, respectively. (e) Line cut along the [0,0,L] direction measured at 3.5 K by integrating over $E=[-0.1, 0.1]$ meV, H~$=[-0.2,0.2]$ and K~$=[-0.2,0.2]$. (f) Line cut along the [H,0,0] direction measured at 3.5 K by integrating over $E=[-0.1, 0.1]$ meV, K~$=[-0.2,0.2]$ and L~$=[0.85,1.15]$.} \label{Diffuse}
\end{figure*}

\subsection{Magnetic phase diagram}

To further study the ordered phase in \TSO, magnetization and specific heat measurements down to 0.1 K were performed. Fig.~\ref{magnetization}(a) is the isothermal magnetization $M(B)$ as a function of the applied field along the $b$ axis with temperature ranging from 0.1 K to 2.5 K. We also checked if there was any slow dynamics persisting in this system, similar with the observation in  DyScO$_3$ \cite{Ke2009,Wu2017,Andriushin2022}. As shown in Fig.~\ref{magnetization}(c) is the measured magnetization at 2 K of~\TSO~with the magnetic field sweep up (red triangles) and down (blue triangles). These two magnetization curves almost overlap with each other, and no hysteresis loop is observed, which is in contrast with the observations in DyScO$_3$ (black circles in Fig.~\ref{magnetization}(c)). These results here suggest that the spin dynamics in~\TSO~is much faster than~\DSO, and it is out of the DC magnetization measurement window. For~\TSO, the zero-field AFM order is gradually suppressed with increasing field, and it finally enters into a polarized ferromagnetic state in high fields. The transitions are evidenced by the sharp peak-like anomalies of the magnetic susceptibility $dM/dB$ at different temperatures, as illustrated in Fig.~\ref{magnetization}(e). The phase boundaries determined by these peak positions of $dM/dB$ are presented in Fig.~\ref{magnetization}(f) (blue diamonds), with the critical field $B_{\mathrm{c}}\simeq0.7$~T at 0.1 K. However, we can see from Fig.~\ref{magnetization}(a), the measured magnetization at $B_{\mathrm{c}}\simeq0.7$~T are still far from the saturation value, even at $T=0.1$~K. It is also noticed in $dM/dB$ curve, the sharp peak is always followed by an additional broad anomaly at the higher fields. As lowering the temperature, we can find that the sharp peak and the broad anomaly smoothly merge into one peak with the width at half maximum around 0.3~T at the base temperature. All these behaviors indicate that there is a field region above $B_{\mathrm{c}}$, where a strong correlation still exists, although the zero-field long-range magnetic order has been suppressed.

Besides the field dependence, the temperature-dependent magnetization $M(T)$ with $B\parallel b$ is presented in Fig.~\ref{magnetization}(b). In small magnetic fields, the $M(T)$ curve first shows a hump-like feature around the AFM transition temperature. This feature is then gradually suppressed to lower temperatures by raising the magnetic field. By taking the temperature derivative (Fig.~\ref{magnetization}(d)), a sharp anomaly is observed in each magnetic susceptibility $dM/dT$ curve with a field range between 0.1~T and 0.5~T, corresponding to the AFM transition points (green triangles in Fig.~\ref{magnetization}(f)). Further increasing the magnetic fields above $B_{\mathrm{c}}$, the sharp anomalies become broader. However, the system does not enter into the fully polarized state in the field region between 0.7~T and 0.9~T. The overall phase diagram becomes straightforward if we plot the phase boundaries extracted from the peak positions of $dM/dB$ and $dM/dT$ curves, together with the contour of the magnetic susceptibility $dM/dB$, as shown in Fig.~\ref{magnetization}(f). Remarkably, the data also reveals the existence of a field region between the $G_xA_y$-type AFM state below 0.7~T and the fully polarized state above 1.0~T.

\begin{figure*}[ht!]
 \includegraphics[width=0.9\textwidth]{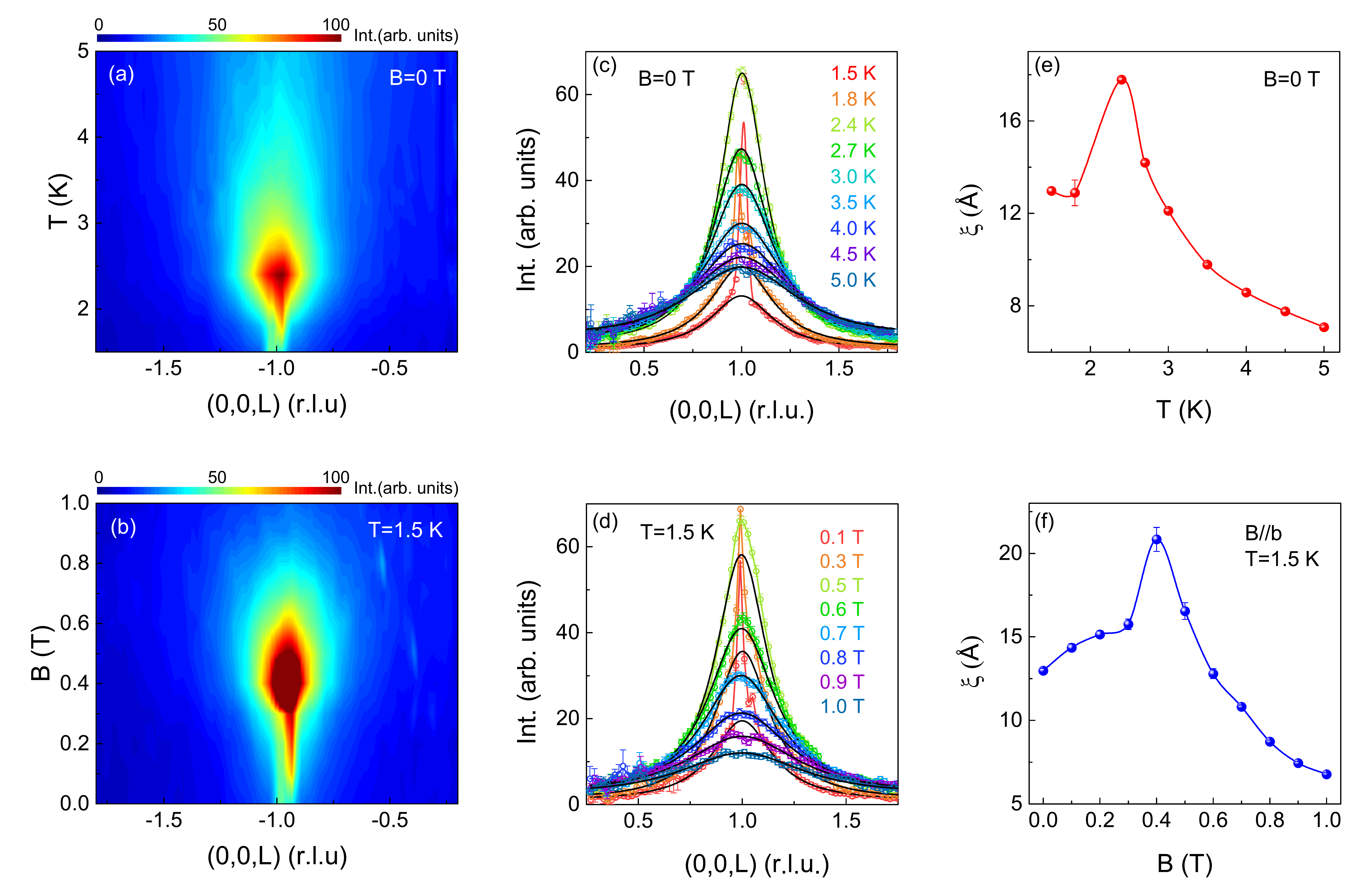}
 \caption{(a) Contour plot of the temperature evolution of the magnetic diffuse scattering along the [0,0,L] direction. (b) Contour plot of the magnetic diffuse scattering field evolution along the [0,0,L] direction. (c) Line cuts along the [0,0,L] direction measured at different temperatures, integrated over $E=[-0.1, 0.1]$ meV, H~$=[-0.2,0.2]$ and K~$=[-0.2,0.2]$. (d) Line cuts along the [0,0,L] direction measured at different fields at 1.5 K, integrated over $E=[-0.1, 0.1]$ meV, H~$=[-0.2,0.2]$ and K~$=[-0.2,0.2]$. The black solid lines are obtained by fitting the lattice Lorentzian function. (e) Temperature dependence of the correlation length along the [0,0,L] direction. (f) Field dependence of the correlation length along the [0,0,L] direction. The correlation length is obtained from the fitting of the lattice Lorentzian function.} \label{diffuse_CL}
\end{figure*}

Specific heat measurements in the field were performed as well to investigate the temperature-field phase diagram. Figure~\ref{HC}(a-c) presents temperature evolution of the magnetic specific heat $C_{\mathrm{M}}/T$ measured with $B \parallel b$. The sharp triangular-like peak in the $C_{\mathrm{M}}/T$ curve is gradually suppressed to lower temperatures with increasing fields (Fig.~\ref{HC}(a)). Further raising the field, this peak turned into a broad anomaly in the intermediate phase as $0.7\leq B \leq 1 $~T [Fig.~\ref{HC}(b)]. Finally, at $B \geq 1$~T, the system enters into the fully polarized state, and a broad Schottky anomaly is observed (Fig.~\ref{HC}(c)).

Here, we also noticed that an upturning feature was observed at the lowest temperatures in a field region between 0.3~T and 1.5~T. The integrated magnetic entropy at different fields is presented in Fig.~\ref{HC}(d-f). The integrated entropy is found to be larger than $R\mathrm{ln}2$ in the region of the intermediate field (Fig.~\ref{HC}(d,e)), where a large upturn is observed in the $C_{\rm \mathrm{M}}/T$ curves. To double-check this phenomenon, field-dependent specific heat at constant temperatures was also measured [Fig.~\ref{HC}(g)]. At higher temperatures (1.2 K $\leq T \leq$ 2.4 K), a sharp peak at the lower field side was observed, indicating the magnetic phase transition. At higher fields, a broad Schottky anomaly was identified, with the peak positions linearly extending to the saturation field with decreasing temperatures. The sharp anomaly and the broad maximum gradually shift towards each other with temperature decreasing and eventually merge into a broad peak at $T=0.8$~K. Further lowering the temperature, the peak becomes even wider, as shown in the inset of Fig.~\ref{HC}(g). The contour plot of the field and temperature-dependent specific heat $C_{\mathrm{M}}/T(T,B)$ is presented in Fig.~\ref{HC}(h). Overlaid on the contour plot are the phase boundaries determined from the magnetization $M(T,B)$ and the specific heat $C_{\mathrm{M}}(T,B)$ measurements. One can see that these two measurements are well consistent with each other. 

The most remarkable phenomenon is the upturning feature of the specific heat at ultra-low temperatures around the intermediate field region. One of the origins may be from the field induced nuclear Schottky anomaly which should contribute a more pronounced effect with increasing magnetic field. On the contrary, we find that the lower temperature upturning is suppressed with further increasing fields above 1.5~T (Fig.~\ref{HC}(c)) in~\TSO.

Moreover, the fields around 1 T should not induce such strong nuclear Schottky anomaly due to the very small nuclear moments. An alternative scenario is that the anomaly may be induced by a much stronger internal field from the ordered f-electrons. To understand these exotic behaviors of specific heat, further studies with neutron scattering or nuclear magnetic resonance (NMR) in this particular field and temperature region are needed.


\begin{table*}[ht!]
\caption{Ground doublet state wave functions of the three iso-structure orthorhombic rare earth perovskites YbAlO$_3$~\cite{Wu2019A,Wu2019B}, DyScO$_3$~\cite{Wu2017}, and TbScO$_3$.}
\centering
\begin{tabular}{p{3cm}<{\raggedright}p{5cm}<{\raggedright}p{3cm}<{\raggedright}p{3cm}<{\raggedright}p{3cm}<{\raggedright}}
    \toprule
               &~~~~~~~~~~~~$|\rm E_{0\pm}\rangle$  &$|\rm\langle E_{0\mp}|S^{\pm}|\rm E_{0\pm}\rangle|$ & $|\rm\langle E_{0\mp}|S^{z}|\rm E_{0\pm}\rangle|$  &  $|\rm\langle E_{0\pm}|S^{z}|\rm E_{0\pm}\rangle|$           \\                  \hline
    ~YbAlO$_3$ &  $\alpha|\pm7/2\rangle+\beta  |\mp5/2\rangle+... $ &$~~~\alpha\beta+...$ &~~~~~~~~0 &$~\alpha^2+\beta^2+...$\\
    ~DyScO$_3$ &  $\alpha|\pm15/2\rangle+\beta  |\pm13/2\rangle+...$ &~~~~~~~~0 &~~~~~~~~0 &$~\alpha^2+\beta^2+...$\\
    ~TbScO$_3$ & $~~1/\sqrt{2}(|-6\rangle \pm  |+6\rangle )$ &~~~~~~~~0 &~~~~~~~~6 &~~~~~~~~0\\

\botrule 
\end{tabular}
\label{tab:CEF}
\end{table*}
\subsection{Magnetic diffuse scattering}
To further explore the spin dynamics, the single crystal neutron scattering of~\TSO~was performed in the (H,0,L) scattering plane at different temperatures and fields. Consistent with a pseudo-doublet ground state, the excited CEF levels are far away from the ground state, and no inelastic magnetic excitations are observed in the low energy window between 0~meV and 2.5~meV at $T=3.5$~K, as presented in Fig.~\ref{Diffuse}(a). Instead, broad magnetic diffuse scattering is identified around $(0,0,\pm1)$ (Fig.~\ref{Diffuse}(b,c)), at 3.5~K, well above the  $T_{\mathrm{N}}=2.53$~K. This indicates that short-range AFM correlations are well developed before the system enters into the long-range ordered state.

With the position-sensitive area detector, we can also access a finite Q-region in the (H,K,1) scattering plane. As shown in Fig.~\ref{Diffuse}(b), this diffuse scattering around $(0,0,1)$ is almost isotropic in the (H,K,1) scattering plane. On the contrary, striped patterns along the [H,0,0] direction is observed in the (H,0,L) scattering plane. For a more quantitative analysis, line cuts along the [0,0,L] and [H,0,0] directions at $T=3.5$~K are made, as shown in Fig.~\ref{Diffuse}(e,f). Here, a lattice Lorentzian function was used to describe the magnetic structure factor~\cite{Igor2015}:
\begin{eqnarray}
S(Q) \propto\frac{\sinh(a/\xi_h)} {\cosh(a/\xi_h)-\cos(\pi{h})} \frac{\sinh(c/\xi_l)}{\cosh(c/\xi_l)-\cos[\pi(l-1)]}
\end{eqnarray}
where $\xi_h$ and $\xi_l$ are the correlation lengths along the $a$ and $c$ axis, respectively. The black solid lines, shown in Fig.~\ref{Diffuse}(e,f), are fittings of the lattice Lorentzian function. The correlation lengths along the $a$ and $c$ axis are about 6.58(10) and 9.79(6)~\AA, respectively. The correlation length along the $c$-axis is larger than that in the $ab$ plane, although the near neighbor distances are similar for these two directions. Furthermore, it is worth noting that the background along the [H,0,0] direction is much larger than that along [0,0,L] direction, suggesting there is stronger diffuse scattering along the [H,0,0] direction. For $T=1.5$ K$ <T_{\mathrm{N}}$, we can also observe the weak striped-like diffuse scattering along the [H,0,0] direction, although a significant part of the intensity is established at $(0,0,\pm1)$ positions due to the formation of long-range magnetic order. This indicates short-range correlation along the [H,0,0] direction in the AFM ordered phase. These are consistent with the quasi-one-dimensional character as a result of the dominant dipole-dipole interaction along the $c$ axis. 

Figure~\ref{diffuse_CL}(a,b) presents the contour plots of the temperature and field-dependent intensity of the (0,0,1) peak. The line cuts of the diffuse scattering along the [0,0,L] direction at different temperatures are illustrated in Fig.~\ref{diffuse_CL}(c). The temperature dependence of the correlation lengths of the diffuse scattering (Fig.~\ref{diffuse_CL}(e)) shows that the correlation length is only about 7.09~\AA\ at $T=5$~K, which is close to the nearest-neighbor Tb-Tb distances ($\approx 7.85$ ~\AA) along the $c$ axis. At the lower temperature, the correlation length increases gradually and reaches a maximum ($\approx 17.78$~\AA) near $T_{\mathrm{N}}$, and then decreases due to the formation of long-range magnetic order. Even at the ordered AFM state, there exists partial fluctuating magnetic moment, which results in the observed weak diffuse scattering with the relatively short correlation length. The field dependence of the magnetic peak at 1.5~K is shown in Fig.~\ref{diffuse_CL}(d). Below the critical field, the magnetic peak is rather sharp due to the long-range order. It is interesting to notice that this magnetic peak did not vanish upon the phase transition. Instead, a very broad peak is observed, which extends up to 1.0~T (Fig.~\ref{diffuse_CL}(b, d)), with the large correlation length $\sim22$~\AA\ found near the phase boundary [Fig.~\ref{diffuse_CL}(f)]. Further studies of this intermediate phase may reveal new physics, such as spin density wave as proposed for the isostructural compound YbAlO$_3$~\cite{Nikitin2021}.


\section{Discussion and Conclusion}
We established the phase diagram and confirmed that strong quasi-one-dimensional spin fluctuations were present in the system of~\TSO. In terms of the crystal structure, local point symmetry, and Ising-like doublet ground states with dominating one-dimensional interaction, there is strong similarity among the three isostructural  perovskite compounds of~\TSO, \DSO~\cite{Wu2017} and \YAO~\cite{Wu2019B}. However they show very different spin dynamics. Broad continuum-like spinon excitations were observed for YbAlO$_3$~\cite{Wu2019A}, while they are absent for DyScO$_3$ and TbScO$_3$. Slow spin dynamics were observed in DyScO$_3$~\cite{Andriushin2022}, while this has not been found for YbAlO$_3$ and TbScO$_3$. What is the underlying mechanism that causes the distinct physics for these three isostructural compounds?

We believe that $the~devil~is~in~the~details$. Listed in the Table~\ref{tab:CEF} are the calculated ground state wave functions for three compounds YbAlO$_3$~\cite{Wu2019B}, DyScO$_3$~\cite{Wu2017} and TbScO$_3$. We found the term $\langle E_{\rm 0\mp}|S^{+}, S^{-}| E_{\rm 0\pm}\rangle =0$ for both DyScO$_3$ and TbScO$_3$. A non-zero contribution of $\langle E_{\rm 0\mp}|S^{+}, S^{-}| E_{\rm 0\pm}\rangle$ was identified for YbAlO$_3$ only. The absence of $\langle E_{\rm 0\mp}|S^{+}, S^{-}| E_{\rm 0\pm}\rangle$ means that the spin Hamiltonian such as $\mathcal{H}^{\rm xy}=J^{\rm xy}S_{\rm i}^{\rm xy}\cdot S_{\rm j}^{\rm xy}$ can neither flip these Ising moments, nor make propagating excitations. Thus, the spinon-like excitations are missing in DyScO$_3$ and TbScO$_3$. This corresponds to an effective spin Hamiltonian only with the Ising term $\mathcal{H}^{\mathrm{z}}=J^{\rm z}S_{\rm i}^{\rm z}\cdot S_{\rm j}^{\rm z}$. In this Ising limit, excited magnetic moments form localized domain walls. In addition, even this flat localized excitation can not be observed in DyScO$_3$ and TbScO$_3$ in neutron scattering experiments, due to the selection rule $\Delta S=1$~\cite{Wu2017}.

On the other hand, for DyScO$_3$, all these terms that connect spin up and down are close to zero ($\langle E_{\rm 0\mp}|S^{\rm \pm}, S^{\rm z}| E_{\rm 0\pm}\rangle=0$). The spin-up moments can relax to the spin-down states by overcoming the large CEF energy barrier~\cite{Guo2018}, and thus significant slow dynamics are observed~\cite{Andriushin2022}. Conversely, Tb is a non-Kramers ion, and the quasi-doublet state consists of two low-lying singlet states. For these singlet states, $\langle E_{\rm 0\pm}|S^{\rm z}| E_{\rm 0\pm}\rangle=0$, but $\langle E_{\rm 0\mp}|S^{\rm z}| E_{\rm 0\pm}\rangle\neq0$. This provides additional channels for tunneling between these singlet states, which may greatly reduce the relaxation time for the spin flipping, and thus, slow spin dynamics are less significant in TbScO$_3$.


In summary, low-temperature magnetic properties of \TSO\ were studied by a combination of CEF calculations, and measurements of magnetization, specific heat, and single-crystal neutron scattering. The results point toward an Ising-like quasi-doublet ground state with wave functions mostly composed of $|\pm 6\rangle$. The Tb$^{3+}$ Ising moments lie in the $ab$ plane with a tilting angle of $\varphi = \pm48^{\mathrm{o}}$ to the $a$ axis, as determined from magnetization measurements. These Ising moments undergo an AFM transition at $T_{\rm N}=2.53$~K, and a $G_xA_y$ magnetic ground state is selected by the dipole-dipole interaction. Similar to \DSO\ and \YAO, the dipolar interaction along the $c$ axis is dominating, which determines quasi-one-dimensional magnetism in~\TSO. We also established the low-temperature phase diagram of~\TSO~with the magnetic field along the easy $b$ axis. Besides the $G_xA_y$ AFM state, a field-induced intermediate state was identified. In this intermediate state, the long-range magnetic order is suppressed but strong one-dimensional correlations still persist. To explore the possible spin configurations in these intermediate field state, further neutron scattering experiments are required. Besides the magnetic phase diagram, we clarified the ground doublet state wave functions of \TSO. The differences in the doublet state wave functions may be responsible for the distinct spin dynamics observed in \YAO\ and \DSO. 

\begin{acknowledgments}
We would like to thank Z. T. Wang for the useful discussions, and G. Davidson for the great support throughout the experiment on Pelican. The authors would also like to acknowledge the beam time awarded by ANSTO through the proposal No.~P9457. The research was supported by the National Key Research and Development Program of China (Grant No.~2021YFA1400400), the National Natural Science Foundation of China (Grants No.~12134020, No.~11974157, No.~12174175, No.~12004426, No.~12005243 and No.~12104255), Shenzhen Fundamental Research Program (Grant No.~JCYJ20220818100405013), the Guangdong Basic and Applied Basic Research Foundation (Grant No.~2021B1515120015 and No.~2022B1515120014), Shenzhen Key Laboratory of Advanced Quantum Functional Materials and Devices (Grant No.~ZDSYS20190902092905285), and the Shenzhen Science and Technology Program (Grant No.~KQTD20200820113047086). The Major Science and Technology Infrastructure Project of Material Genome Big-science Facilities Platform supported by Municipal Development and Reform Commission of Shenzhen. A.P. acknowledges the support by the Scientific User Facilities Division, Office of Basic Energy Sciences, US Department of Energy.


\end{acknowledgments}


\begin{thebibliography}{10}
\bibitem{Kimura2003}
T. Kimura, T. Goto, H. Shintani, K. Ishizaka, T. Arima, Y. Tokura, Magnetic control of ferroelectric polarization, \href{https://www.nature.com/articles/nature02018}{Nature 426, 55 (2003)}.

\bibitem{Goto2004}
T. Goto, T. Kimura, G. Lawes, A. P. Ramirez, Y. Tokura, Ferroelectricity and giant magnetocapacitance in perovskite rare-earth manganites, \href{https://journals.aps.org/prl/abstract/10.1103/PhysRevLett.92.257201}{Phys. Rev. Lett. 92, 257201 (2004)}.

\bibitem{Kimel2004}
A. V. Kimel, A. Kirilyuk, A. Tsvetkov, R. V. Pisarev, Th. Rasing, Laser-induced ultrafast spin reorientation in the antiferromagnet TmFeO$_3$, \href{https://www.nature.com/articles/nature02659}{Nature 429, 850 (2004)}.

\bibitem{Kimel2005}
A. V. Kimel, A. Kirilyuk, P. A. Usachev, R. V. Pisarev, A. M. Balbashov, Th. Rasing, Ultrafast non-thermal control of magnetization by instantaneous photomagnetic pulses, \href{https://www.nature.com/articles/nature03564}{Nature 435, 655 (2005)}.

\bibitem{Wuyaodong2019}
Y. D. Wu, Y. L. Qin, X. H. Ma, R. W. Li, Y. Y. Wei, Z. F. Zi, Large rotating magnetocaloric effect at low magnetic fields in the Ising-like antiferromagnet DyScO$_3$ single crystal, \href{https://doi.org/10.1016/j.jallcom.2018.11.047}{J. Alloys Compd. 777, 673 (2019)}.

\bibitem{Jia2019}
J. H. Jia, Y. J. Ke, X. X. Zhang, J. F. Wang, L. Su, Y. D. Wu, Z. C. Xia, Giant magnetocaloric effect in the antiferromagnet GdScO$_3$ single crystal, \href{https://doi.org/10.1016/j.jallcom.2019.06.361}{J. Alloys Compd. 803, 992 (2019)}.






\bibitem{Nikitin2018}
S. E. Nikitin, L. S. Wu, A. S. Sefat, K. A. Shaykhutdinov, Z. Lu, S. Meng, E. V. Pomjakushina, K. Conder, G. Ehlers, M. D. Lumsden, et.al., Decoupled Spin Dynamics in the Rare-Earth Orthoferrite YbFeO$_{3}$: Evolution of Magnetic Excitations through the Spin-Reorientation Transition, \href{https://doi.org/10.1103/PhysRevB.98.064424}{Phys. Rev. B 98, 064424 (2018)}.

\bibitem{Wu2019A}
L. S. Wu, S. E. Nikitin, Z. Wang, W. Zhu, C. D. Batista, A. M. Tsvelik, A. M. Samarakoon, D. A. Tennant, M. Brando, L. Vasylechko, et. al., Tomonaga–Luttinger Liquid Behavior and Spinon Confinement in YbAlO$_{3}$, \href{https://www.nature.com/articles/s41467-019-08485-7}{Nat. Commun. 10, 698 (2019)}.

\bibitem{Podlesnyak2021}
A. Podlesnyak, S. E. Nikitin, and G. Ehlers, Low-energy spin dynamics in rare-earth perovskite oxides, \href{https://10.1088/1361-648X/ac1367}{J. Phys. Condens. Matter 33, 403001 (2021)}.

\bibitem{Gorodetsky1968}
G. Gorodetsky, B. Sharon, S. Shtrikman, Magnetic Properties of an Antiferromagnetic Orthoferrite, \href{https://doi.org/10.1063/1.1656309}{J. Appl. Phys. 39, 1371 (1968)}.

\bibitem{Belov1976}
K. P. Belov, A. K. Zvezdin, A. M. Kadomtseva, R. Z. Levitin, Spin-reorientation transitions in rare-earth magnets, \href{https://iopscience.iop.org/article/10.1070/PU1976v019n07ABEH005274/meta}{Sov. Phys. Usp. 19, 574 (1976)}.


\bibitem{Yamaguchi1973}
T. Yamaguchi, K. Tsushima, Magnetic symmetry of rare-earth orthochromites and orthoferrites, \href{https://doi.org/10.1103/PhysRevB.8.5187}{Phys. Rev. B 8, 5187 (1973)}.

\bibitem{Shapiro1974}
S. M. Shapiro, J. D. Axe, J. P. Remeika, Neutron scattering studies of spin waves in rare-earth orthoferrites, \href{https://doi.org/10.1103/PhysRevB.10.2014}{Phys. Rev. B 10 2014 (1974)}.

\bibitem{Hahn2014}
S. E. Hahn, A. A. Podlesnyak, G. Ehlers, G. E. Granroth, R. S. Fishman, A. I. Kolesnikov, E. Pomjakushina, K. Conder, Inelastic neutron scattering studies of YFeO$_3$, \href{https://doi.org/10.1103/PhysRevB.89.014420}{Phys. Rev. B 89, 014420 (2014)}.



\bibitem{White1969}
R. L. White, Review of Recent Work on the Magnetic and Spectroscopic Properties of the Rare-Earth Orthoferrites, \href{https://doi.org/10.1063/1.1657530}{J. Appl. Phys. 40, 1061 (1969)}.

\bibitem{Yamaguchi1974}
T. Yamaguchi, Theory of spin reorientation in rare-earth orthochromites and orthoferrites, \href{https://doi.org/10.1016/S0022-3697(74)80003-X}{J. Phys. Chem. Solids 35, 479 (1974)}.

\bibitem{Artyukhin2012}
S. Artyukhin,  M. Mostovoy, N. P. Jensen, Duc Le, K. Prokes, V. G. de Paula, H. N. Bordallo, A. Maljuk, S. Landsgesell, H. Ryll, et. al., Solitonic Lattice and Yukawa Forces in the Rare-Earth Orthoferrite TbFeO$_{3}$, \href{https://www.nature.com/articles/nmat3358}{Nat. Mater. 11, 694 (2012)}.

\bibitem{Tokunaga2008}
Y. Tokunaga, S. Iguchi, T. Arima, and Y. Tokura, Magnetic-Field-Induced Ferroelectric State in DyFeO$_{3}$, \href{https://doi.org/10.1103/PhysRevLett.101.097205}{Phys. Rev. Lett. 101, 097205 (2008)}.


\bibitem{Sheng2020}
Jie-Ming Sheng, Xu-Cai Kan, Han Ge, Pei-Qian Yuan, Lei Zhang, Nan Zhao, Zong-Mei Song, Yuan-Yin Yao, Ji-Ning Tang, Shan-Min Wang, Low Temperature Magnetism in the Rare-Earth Perovskite GdScO$_{3}$, \href{https://iopscience.iop.org/article/10.1088/1674-1056/ab8200/meta}{Chinese Physics B 29, 057503 (2020)}.


\bibitem{Wu2017}
L. S. Wu, S. E. Nikitin, M. Frontzek, A. I. Kolesnikov, G. Ehlers, M. D. Lumsden, K. A. Shaykhutdinov, E.-J. Guo, A. T. Savici, Z. Gai, A. S. Sefat, and A. Podlesnyak, Magnetic Ground State of the Ising-like Antiferromagnet DyScO$_{3}$, \href{https://doi.org/10.1103/PhysRevB.96.144407}{Phys. Rev. B 96, 144407 (2017)}.

\bibitem{Wu2019B}
L. S. Wu, S. E. Nikitin, M. Brando, L. Vasylechko, G. Ehlers, M. Frontzek, A. T. Savici, G. Sala, A. D. Christianson, M. D. Lumsden, and A. Podlesnyak, Antiferromagnetic Ordering and Dipolar Interactions of YbAlO$_{3}$, \href{https://doi.org/10.1103/PhysRevB.99.195117}{Phys. Rev. B 99, 195117 (2019)}.

\bibitem{Schlom2007}
D. G. Schlom, L. Q. Chen, C. B. Eom, K. M. Rabe, S. K. Streiffer, J. M. Triscone, Strain tuning of ferroelectric thin films, \href{https://www.annualreviews.org/doi/abs/10.1146/annurev.matsci.37.061206.113016}{Ann. Rev. Mater. Res. 37, 589 (2007)}.

\bibitem{Haeni2004}
J. H. Haeni, P. Irvin, W. Chang, R. Uecker, P. Reiche, Y. L. Li, S. Choudhury, W. Tian, M. E. Hawley, B. Craigo, A. K. Tagantsev, X. Q. Pan, S. K. Streiffer, L. Q. Chen, S. W. Kirchoefer, J. Levy, and D. G. Schlom, Room-Temperature Ferroelectricity in Strained SrTiO$_{3}$, \href{https://www.nature.com/articles/nature02773}{Nature 430, 758 (2004)}.

\bibitem{Lustikova2022}
J. Lustikova, R.-F. Wang, Y. Zhong, S. Wang, A. Kumatani, X.-C. Ma, Q.-K. Xue, and Y. P. Chen, Magnetotransport of Thin Film Sr$_{1-x}$La$_{x}$CuO$_{2}$ on (110) DyScO$_{3}$, \href{https://iopscience.iop.org/article/10.35848/1347-4065/ac50bc/meta}{Jpn. J. Appl. Phys. 61, 040904 (2022)}.


\bibitem{Uecker2008}
R. Uecker, B. Velickov, D. Klimm, R. Bertram, M. Bernhagen, M. Rabe, M. Albrecht, R. Fornari, and D. G. Schlom, Properties of Rare-Earth Scandate Single Crystals (Re=Nd-Dy), \href{https://www.sciencedirect.com/science/article/pii/S0022024808000675}{J. Cryst. Growth 310, 2649 (2008)}.

\bibitem{Hallsensor}
Magnetometry by means of Hall micro-probes in the Quantum Design PPMS, \emph{Quantum Design Application Note} 1084-701.

\bibitem{Cavallini2004}
A. Cavallini, B. Fraboni, F. Capotondi, L. Sorba, G. Biasiol, Deep Levels in MBE Grown AlGaAs$/$GaAs Heterostructures, \href{https://www.sciencedirect.com/science/article/abs/pii/S0167931704002503}{Microelectron. Eng., 73-74, 954 (2004)}.

\bibitem{Candini2006}
A. Candini, G. C. Gazzadi, A. di Bona, M. Affronte, D. Ercolani, G. Biasiol, and L. Sorba, Hall Nano-Probes Fabricated by Focused Ion Beam, \href{https://iopscience.iop.org/article/10.1088/0957-4484/17/9/005/meta}{Nanotechnology, 17, 2105 (2006)}.

\bibitem{Yu2013}
D. Yu, R. Mole, T. Noakes, S. Kennedy, and R. Robinson, Pelican - a Time of Flight Cold Neutron Polarization Analysis Spectrometer at OPAL, \href{https://journals.jps.jp/doi/abs/10.7566/JPSJS.82SA.SA027}{J. Phys. Soc. Jpn. 82, SA027 (2013)}.

\bibitem{Wang}
J. Liu, B. Liu, L. Yuan, B. Li, L. Xie, X. Chen, H. Zhang, D. Xu, W. Tong, J. Wang, et al., Evidence for a Gapless Dirac Spin-Liquid Ground State in a Spin-3/2 Triangular-Lattice Antiferromagnet, \href{https://iopscience.iop.org/article/10.1088/1367-2630/abe813/meta}{New J. Phys. 23, 033040 (2021)}.

\bibitem{Azuah2009}
R. T. Azuah, L. R. Kneller, Y. Qiu, P. L. W. Tregenna-Piggott, C. M. Brown, J. R. D. Copley, and R. M. Dimeo, DAVE: A Comprehensive Software Suite for the Reduction, Visualization, and Analysis of Low Energy Neutron Spectroscopic Data, \href{https://www.ncbi.nlm.nih.gov/pmc/articles/PMC4646530/}{J. Res. Natl. Inst. Stand. Technol. 114, 341 (2009)}.

\bibitem{Ewings2016}
R.A. Ewings, A. Buts, M.D. Le, J. van Duijn, I. Bustinduy, T.G. Perring,
Horace: Software for the analysis of data from single crystal spectroscopy experiments at time-of-flight neutron instruments, Nuclear Instruments and Methods in Physics Research Section A: Accelerators, Spectrometers, Detectors and Associated Equipment, \href{https://doi.org/10.1016/j.nima.2016.07.036}{Nuclear Instruments and Methods in Physics Research Section A: Accelerators, Spectrometers, Detectors and Associated Equipment 834, 132-142 (2016)}.


\bibitem{MCPHASE2004}
M. Rotter, Using McPhase to calculate magnetic phase diagrams of rare earth compounds, \href{https://www.sciencedirect.com/science/article/abs/pii/S0304885303022881}{J. Magn. Magn. Mater., 272-276, E481 (2004)}.

\bibitem{Valiev2021}
U. V. Valiev, D. N. Karimov, M. G. Brik, C. G. Ma, R. R. Vildanov, F. K. Turotov, and V. O. Pelenovich, Zeeman Splitting Features of Electronic States of Rare Earth Ions in TbF$_{3}$ Crystal, \href{https://www.sciencedirect.com/science/article/abs/pii/S0925346721003426}{Opt. Mater. 117, 111141 (2021)}.

\bibitem{Holmes1971}
L. Holmes and H. J. Guggenheim, Ferromagnetism in TbF$_{3}$, \href{https://hal.archives-ouvertes.fr/jpa-00213987}{J. Phys. Colloques 32, C1 (1971)}.

\bibitem{Holmes1968}
L. Holmes, R. Sherwood, L. G. Van Uitert, Metamagnetism in TbAlO$_{3}$, \href{https://aip.scitation.org/doi/abs/10.1063/1.1656310}{J. Appl. Phys. 39, 1373 (1968)}.

\bibitem{Hufner1967}
S. H{\"u}fner, L. Holmes, F. Varsanyi, L. G. Van Uitert, Optical absorption spectra of antiferromagnetic TbAlO$_{3}$, \href{https://www.sciencedirect.com/science/article/abs/pii/0375960167906615}{Phys. Lett. A 25, 301 (1967)}.


\bibitem{Ke2009}
X. Ke, C. Adamo, D. G. Schlom, M. Bernhagen, R. Uecker, and P. Schiffer, Low temperature magnetism in the perovskite substrate DyScO$_{3}$, \href{https://aip.scitation.org/doi/full/10.1063/1.3117190}{Applied Physics Letters 94, 152503 (2009)}.

\bibitem{Andriushin2022}
N. D. Andriushin, S. E. Nikitin, G. Ehlers, and A. Podlesnyak, Slow Spin Dynamics and Quantum Tunneling of Magnetization in the Dipolar Antiferromagnet DyScO$_{3}$, \href{https://doi.org/10.1103/PhysRevB.106.104427}{Phys. Rev. B 106, 104427 (2022)}.


\bibitem{Igor2015}
I. Zaliznyak, A. T. Savici, M. Lumsden, A. Tsvelik, R. Hu, and C. Petrovic, Spin-liquid polymorphism in a correlated electron system on the threshold of superconductivity, \href{https://doi.org/10.1073/pnas.1503559112}{Proc. Natl. Acad. Sci. 112, 10316–10320 (2015)}.

\bibitem{Nikitin2021}
S. E. Nikitin, S. Nishimoto, Y. Fan, J. Wu, L. S. Wu, A. S. Sukhanov, M. Brando, N. S. Pavlovskii, J. Xu, L. Vasylechko, R. Yu, and A. Podlesnyak, Multiple Fermion Scattering in the Weakly Coupled Spin-Chain Compound YbAlO$_{3}$, \href{https://www.nature.com/articles/s41467-021-23585-z}{Nat. Commun. 12, 3599 (2021)}.

\bibitem{Guo2018}
F. S. Guo, B. M. Day, Y. C. Chen, M. L. Tong, A. Mansikkam{\"a}ki, R. A. Layfield, Magnetic hysteresis up to 80 kelvin in a dysprosium metallocene single-molecule magnet, \href{https://www.science.org/doi/10.1126/science.aav0652}{Science 362, 1400 (2018)}.




\end{thebibliography}
\end{document}